**Title**

Low power reconfigurability and reduced crosstalk in integrated photonic circuits fabricated by femtosecond laser micromachining

**Authors**

*Francesco Ceccarelli**

Istituto di Fotonica e Nanotecnologie – Consiglio Nazionale delle Ricerche (IFN-CNR), piazza L. da Vinci 32, 20133 Milano, Italy.

Dipartimento di Fisica - Politecnico di Milano, piazza L. da Vinci 32, 20133 Milano, Italy.

Email: francesco.ceccarelli@polimi.it

Telephone: +39 2399 6124

*Simone Atzeni*

Istituto di Fotonica e Nanotecnologie – Consiglio Nazionale delle Ricerche (IFN-CNR), piazza L. da Vinci 32, 20133 Milano, Italy.

Dipartimento di Fisica - Politecnico di Milano, piazza L. da Vinci 32, 20133 Milano, Italy.

Email: simone.atzeni@polimi.it

Telephone: +39 2399 6584

*Ciro Pentangelo*

Istituto di Fotonica e Nanotecnologie – Consiglio Nazionale delle Ricerche (IFN-CNR), piazza L. da Vinci 32, 20133 Milano, Italy.

Dipartimento di Fisica - Politecnico di Milano, piazza L. da Vinci 32, 20133 Milano, Italy.

Email: ciro.pentangelo@mail.polimi.it

Telephone: +39 2399 6583

*Francesco Pellegatta*

Istituto di Fotonica e Nanotecnologie – Consiglio Nazionale delle Ricerche (IFN-CNR), piazza L. da Vinci 32, 20133 Milano, Italy.

Dipartimento di Fisica - Politecnico di Milano, piazza L. da Vinci 32, 20133 Milano, Italy.

Email: francesco1.pellegatta@mail.polimi.it

Telephone: +39 2399 6583

*Andrea Crespi*

Istituto di Fotonica e Nanotecnologie – Consiglio Nazionale delle Ricerche (IFN-CNR), piazza L. da Vinci 32, 20133 Milano, Italy.

Dipartimento di Fisica - Politecnico di Milano, piazza L. da Vinci 32, 20133 Milano, Italy.

Email: andrea.crespi@polimi.it

Telephone: +39 2399 6587

*Roberto Osellame*

Istituto di Fotonica e Nanotecnologie – Consiglio Nazionale delle Ricerche (IFN-CNR), piazza L. da Vinci 32, 20133 Milano, Italy.

Dipartimento di Fisica - Politecnico di Milano, piazza L. da Vinci 32, 20133 Milano, Italy.

Email: roberto.osellame@polimi.it

Telephone: +39 2399 6075

**Abstract**

Femtosecond laser writing is a powerful technique that allows rapid and cost-effective fabrication of photonic integrated circuits with unique three-dimensional geometries. In particular, the possibility to reconfigure such devices by thermo-optic phase shifters represents a paramount feature, exploited to produce adaptive and programmable circuits. However, the scalability is strongly limited by the flaws of current thermal phase shifters, which require hundreds of milliwatts to operate and exhibit large thermal crosstalk. In this work, thermally-insulating three-dimensional microstructures are exploited to decrease the power needed to induce a $2\pi$ phase shift down to 37 mW and to reduce the crosstalk to a few percent. Further improvement is demonstrated when operating in vacuum, with sub-milliwatt power dissipation and negligible crosstalk. These results pave the way towards the demonstration of complex programmable integrated photonic circuits fabricated by femtosecond laser writing, thus opening exciting perspectives in integrated quantum photonics.

1. Introduction

Integrated photonics is considered today an enabling technology capable of providing a level of miniaturization and complexity hardly attainable with bulk optical components and thus attracting a lot of attention from photonic quantum information processing (QIP)[1]. In addition, excellent phase stability is surely among the qualities that photonic integrated circuits (PICs) can claim and several physical effects[2], [3], [4] can be exploited to obtain controlled and dynamically adjusted phase shifts between optical signals. However, many applications in QIP actually require only a quasi-static reconfiguration of the circuit operation, either to finely tune a few circuit parameters[5] or to completely change the implemented unitary transformation[6]. To this purpose, a commonly exploited approach is to harness the dependence of the refractive index on the local temperature: the thermo-optic effect. Thermal phase shifting has a straightforward implementation, because it requires only the integration of an electrical microheater (i.e. usually a resistor that dissipates electrical power by Joule effect), it does not introduce additional photon losses and, at the same time, it provides an excellent performance in terms of stability and accuracy.

Silicon on insulator (SOI)[7], [8] is a PIC platform that represents today the gold standard in terms of miniaturization, density of components and scaling to mass production. However, at present, several other PIC technologies are also adopted in the most advanced QIP experiments. These include silica on silicon[9], silicon nitride[10], lithium niobate[11], UV laser written circuits[12] and femtosecond laser writing (FLW)[13]. In particular, FLW of silicate glasses is a fabrication platform which intrinsically inherits all the advantages of the silica-based technologies (i.e. wide transparency wavelength range and efficient coupling with standard optical fibers and off-the-shelf optical components) and, at the same time, alleviates the strict constraints of the planar photolithographic process, as demonstrated by the results published in the last decade[14], [15]. As a matter of fact, FLW does not require a photolithographic mask, thus allowing for rapid and cost-effective fabrication of photonic circuits with arbitrary topology[16] and, since the laser induces a modification confined in the focal volume of the beam, this technique enables the direct inscription of circuits featuring three-dimensional (3D) waveguide geometries that would be unfeasible with a planar process[17]. Furthermore, femtosecond laser written PICs (FLW-PICs) allows one to fabricate low birefringence waveguides (down to 1.2E-6[18]), which are mandatory for QIP applications in which the information is encoded in the polarization state of single photons. This feature goes together with low propagation losses (less than 0.3 dB/cm at 1550 nm[18]), which are essential to scale the number of single photons processable in a QIP experiment. Lastly, FLW can be employed to process a wide range of materials beyond silicate glasses, enabling the investigation of integrated quantum components like sources[19] or memories[20] in order to produce hybrid integrated circuits able to exploit the best performance of each component required by QIP[21], [22].

The introduction of thermal phase shifters in FLW-PICs has greatly extended both the quality and the applicability of these devices[23], [24], [25]. However, some design challenges prevent their deployment on a larger scale. Firstly, thermal shifters in FLW-PICs suffer from high power consumption and, up to now, a $2\pi$ phase modulation has been demonstrated only with a dissipation of hundreds of milliwatts[4], [26]. Since the total power dissipation that a photonic circuit can tolerate with no active cooling is limited to few watts, the total number of thermal shifters that can be integrated in the same chip is usually limited to no more than a dozen[26]. Another obstacle is represented by the thermal crosstalk: indeed, when a thermal shifter dissipates power, the heat can also reach waveguides different from the target one, inducing on them an undesired phase shift. To achieve an accurate phase control when multiple phases are tuned simultaneously, a massive calibration procedure is required including all possible combinations of dissipated powers in the different shifters, to also take into account that thermal shifters may have a nonlinear response with temperature[27]. This also means that all thermal shifters will have to be adjusted even if only one phase needs to be varied. Moreover, the additive nature of the crosstalk may require to set the waveguide temperature for a given phase shift at a much higher value than that required when operating the thermal shifters individually, thus jeopardizing both the stability of the phase response and the reliability of the microheater. Since the thermal crosstalk depends on the distance between thermal shifter and waveguides, a reduction of this phenomenon is necessary to increase the integration density of reconfigurable FLW-PICs that, up to now, have never been reported with an inter-waveguide pitch lower than 100 μm[26]. In order to cope with these challenges, some solutions have been already proposed in the literature. On the one hand, in a previous work[27] we have proved that the power dissipation can be moderately reduced with a compact design of the microheater and, thanks to this approach, we have demonstrated a $2\pi$ phase shift for light at 800 nm by dissipating only 200 mW. On the other hand, Chaboyer et al.[28] have achieved a similar level of power dissipation, along with some reduction of the thermal crosstalk, by introducing an isolation trench between the two arms of a thermally-reconfigurable Mach-Zehnder interferometer (MZI).

In this work, by 3D structuring the glass we are able to significantly concentrate the heat diffusion around the waveguide on which we want to induce the phase shift. This novel approach enables orders-of-magnitude reduction of the dissipated power and of the thermal crosstalk, allowing us to present a new generation of thermal phase shifters in glass photonic circuits with great potential for scalability and compactness. Power dissipation, dynamic response, thermal crosstalk and stability of the phase response are thoroughly characterized both in air and in vacuum, where we achieve an unprecedented level of performance.

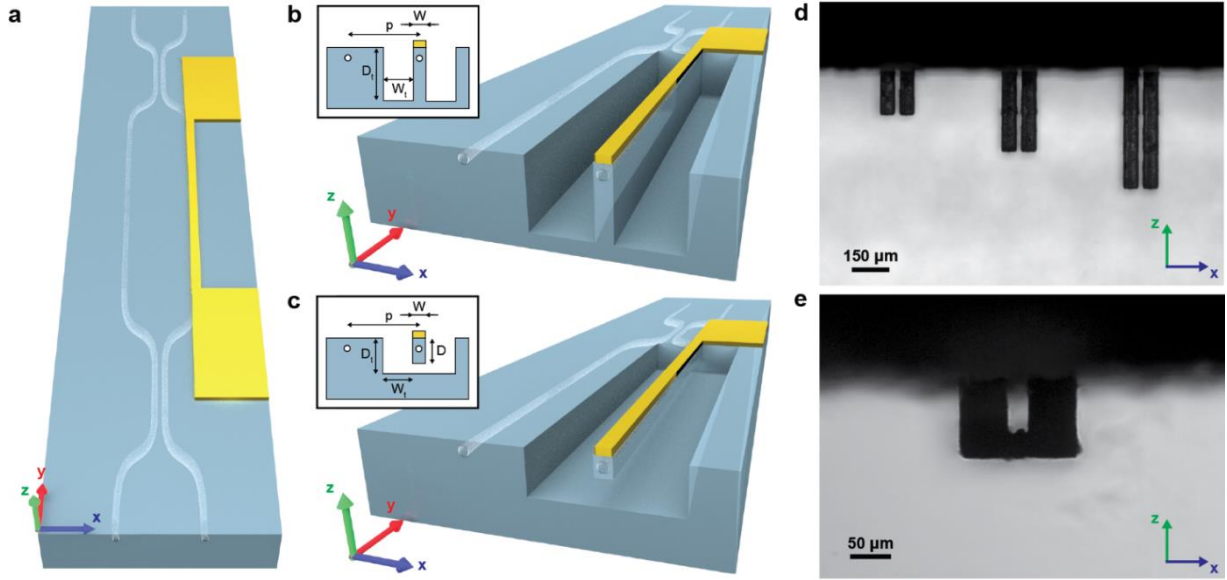

*Figure 1. The reconfigurable MZIs used to demonstrate the capabilities of the new technological platform. (a) Basis structure of the device with no isolation structures. (b) Cross section of the device with deep isolation trenches. (c) Cross section of the device with the bridge waveguide. (d) Photomicrograph of the isolation trenches ($D_t$ = 150, 300, 450 μm) seen through the side of the substrate. (e) Photomicrograph of the bridge waveguide seen through the side of the substrate.*

## 2. Device structure and operation

Without loss of generality, the discussion will be restricted to reconfigurable MZIs, which represent the basic building blocks for a universal multiport device able to implement any unitary transformation of a photonic quantum state[29]. The interferometers are inscribed in a boro-aluminosilicate glass (Corning EAGLE XG[30], 1.1 mm thick) and optimized for single-mode operation at 1550 nm wavelength. The basic structure of the MZI is depicted in Figure 1a: the optical waveguides are inscribed at 30 μm below the chip surface, forming an optical circuit composed of two 3 dB directional couplers (interaction length 1.2 mm, coupling distance 9 μm), that are connected by sinusoidal S-bend waveguides (minimum radius of curvature 45 mm) to the central arms of the interferometer (two straight waveguides having length *L* and distance *p*). Light propagation is characterized by losses of 0.29 dB/cm and a mode diameter ($1/e^2$) of 9 μm. Such a circuit enables the coupling to standard single-mode optical fibers (SMF-28) with losses as low as 0.27 dB/facet.

Given coherent light injected in one of the two input ports, the optical power $I_{out}$ measured at one of the two outputs can be modulated by acting on the phase difference $\varphi$ that is present between the two optical paths. Mathematically speaking

$$I_{out} = \frac{I_{max}}{2}[1 + v \cos \varphi], \quad (1)$$

where $I_{max}$ is the sum of the two output optical powers and $v$ is the visibility of the interference fringe. In order to tune the phase difference $\varphi$, we exploit a Cr-Au resistive microheater (length $L_r = L$ and width $W_r$) fabricated on top of one of the two central arms (see Figure 1a). Given the linearity of the heat equation and assuming a linear relation between refractive index and temperature change of the substrate, we can model the phase difference $\varphi$ between the two arms as[4]

$$\varphi = \Phi + \alpha P, \quad (2)$$

where $\Phi$ is the phase difference when the microheater is not biased, $\alpha$ is the tuning coefficient of the interferometer and $P$ is the thermal power emitted by the resistive microheater. This is equivalent to the electrical power dissipated by Joule effect on such resistance, which in our experiments was retrieved by electrical measurements using the known relation $P = V I$, where $V$ is the voltage drop across the resistor and $I$ is the current passing in it. In order to induce a given phase difference with minimal power dissipation, it is necessary to maximize the tuning coefficient $\alpha$. To this aim, it is useful to derive an analytical expression for this parameter. Considering an infinitesimal waveguide segment having length $dl$, the corresponding phase $d\varphi$ induced by the microheater is

$$d\varphi = \frac{2\pi}{\lambda} \Delta n(l)\, dl = \frac{2\pi n_t}{\lambda} \Delta T(l)\, dl, \quad (3)$$

where $\lambda$ is the wavelength, $\Delta n(l)$ and $\Delta T(l)$ are, respectively, the refractive index and the temperature difference between the two waveguides at a given coordinate $l$ and $n_t$ is the thermo-optic coefficient of the substrate. By integrating Equation 3 over the entire optical path $\gamma$ and substituting $\Delta\varphi = \varphi - \Phi$ into Equation 2, one reads

$$\alpha = \frac{\Delta\varphi}{P} = \frac{2\pi n_t}{\lambda}\int_\gamma \frac{\Delta T(l)}{P} dl \cong \frac{2\pi n_t}{\lambda}\frac{\Delta T}{\overline{P}} = \frac{2\pi n_t}{\lambda}\overline{R}, \qquad (4)$$

where we have introduced the linear power dissipation density $\overline{P} = P/L$ and the thermal efficiency factor $\overline{R} = \Delta T/\overline{P}$. The approximation in Equation 4 is valid whenever $L$ is much greater than the transverse dimensions of the device. Indeed, in this condition the thermal problem can be studied with a two-dimensional (2D) geometry, in which the heat flux is always orthogonal to the light propagation and in which there exists a temperature difference $\Delta T$ that is constant and different from zero only for a length $L$, as long as the waveguide is beneath the microheater. The quantity $\overline{R}$ has the dimensions of a 2D thermal resistance and depends only on the transverse geometry of the device and on the thermal conductivity $k$ of the substrate. For given material and wavelength, $\overline{R}$ represents the only degree of freedom for the optimization of $\alpha$. In particular, this factor can be increased by avoiding the heat diffusion all over the chip and, thus, by thermally isolating the target waveguide from the rest of the circuit. Of course, this approach is beneficial also for the thermal crosstalk between different integrated interferometers in the same substrate.

Two different isolating structures are investigated in this work: deep isolation trenches (Figure 1b) and the bridge waveguide (Figure 1c). On the one hand, the isolation trenches (Figure 1b) are fabricated by removing glass boxes (dimensions $L_t \times W_t \times D_t$) from both sides of the microheater. They are removed as close as possible to the waveguide, in order to limit the portion of glass subject to the heating and to reduce the width $W$ of the slab that thermally connects the waveguide to the rest of the circuit. A nominal width $W = 20$ μm represents a value that, given the dimensions of the optical mode, does not affect the insertion losses of the circuit and, at the same time, is compatible with the fabrication of a microheater that occupies all the area between the two trenches (i.e. $W_r = W$). The dimensions of the boxes are chosen to match the length of the microheater (i.e. $L_t = L_r = L$) and to allow the reduction of the inter-waveguide pitch $p$ down to 80 μm (i.e. $W_t = p - W = 60$ μm). A photomicrograph of deep isolation trenches having $D_t$ = 150, 300, 450 μm is reported in Figure 1d. On the other hand, the bridge waveguide (Figure 1c) is a structure based on the former one, but in which the glass is removed also under the optical path, thus leaving the waveguide inside a suspended bridge whose section has nominal dimensions $W = 20$ μm and $D = 60$ μm. The dimensions of the lateral boxes are $L_t = L_r = L$, $W_t = 60$ μm and $D_t = 90$ μm. A photomicrograph of a bridge waveguide is reported in Figure 1e.

### 3. Experimental results

To show the capabilities of the new technology, we report here the results of the experimental characterization performed on the photonic circuits described in Section 2. A detailed description of fabrication processes, experimental setups and simulation techniques is reported in the Supporting Information.

#### 3.1. Power dissipation

The electrical power $P_{2\pi}$ that a microheater must dissipate to induce the maximum useful phase shift (i.e. $\Delta\varphi = 2\pi$) is directly related to the tuning coefficient $\alpha$ through the relation

$$P_{2\pi} = \frac{2\pi}{\alpha} = \frac{\lambda}{n_t \bar{R}}. \qquad (5)$$

This quantity has been experimentally characterized on reconfigurable MZIs ($L = 3$ mm, $p = 127$ μm) featuring isolation trenches with depth $D_t$ ranging from 0 (no trenches) to 450 μm. The power dissipation $P_{2\pi}$ is reported in Figure 2a as a function of the depth $D_t$: the use of isolation trenches allows a reduction of the power dissipation of more than an order of magnitude, from 611 mW for a MZI with no isolation, to 57 mW for a MZI with trenches 450 μm deep. It is clear from Figure 2a that, for $D_t > 300$ μm, the power dissipation has already saturated to the minimum value and a further increase of the thermal isolation produces no effect on the MZI performance. However, modelling such phenomenon by considering heat dissipation only through the glass slab underneath the microheater does not predict this saturation trend. Indeed, by considering a trench depth $D_t$ sufficiently large, the 2D thermal resistance $\overline{R_{slab}}$ (see the inset of Figure 2a) between the microheater and the heat sink at the bottom of the substrate (fixed at room temperature $T_{room}$) can be calculated as

$$\overline{R_{slab}} = \frac{D_t}{kW}. \qquad (6)$$

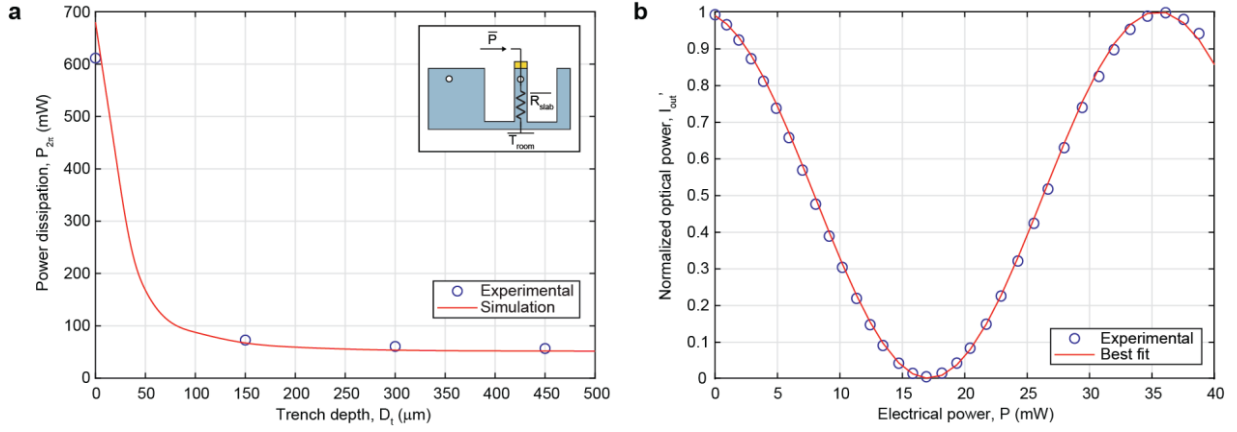

***Figure 2.*** Power dissipation necessary to induce a complete phase shift on a MZI. (a) Power dissipation $P_{2\pi}$ as a function of the trench depth $D_t$. The simulation line is achieved by considering thermal leakage through the air gaps. Inset: lumped-element model of the 2D thermal problem assuming no conduction through the air gaps. (b) Normalized optical power $I_{out}'$ as a function of the electrical power $P$ dissipated by a microheater on a bridge waveguide.

This quantity determines the temperature of the target waveguide, that is $T_1 = T_{room} + \overline{R_{slab}}\,\overline{P}$. By assuming negligible heating on the second arm of the MZI (i.e. $T_2 = T_{room}$), the temperature difference between the waveguides is $\Delta T = \overline{R_{slab}}\,\overline{P}$ and, therefore, the thermal efficiency factor is $\overline{R} = \overline{R_{slab}}$. In conclusion, Equation 5 becomes

$$P_{2\pi} = \frac{\lambda kW}{n_t D_t}. \qquad (7)$$

It is clear from Equation 7 that, for an increasing trench depth $D_t$, the power dissipation $P_{2\pi}$ should approach zero. On the contrary, the saturation observed during the experimental characterization is consistent with the presence of a thermal leakage that breaks the isolation achieved with the trenches. Indeed, finite-element simulations demonstrate that, by taking into account the thermal conduction through the air gaps, the experimental behavior can be predicted with an error that is lower than 12% over the entire dataset (see Figure 2a).

Even though the presence of air limits the effectiveness of the isolation structures, a further reduction of the power dissipation $P_{2\pi}$ can be achieved with the bridge waveguide. Figure 2b reports the normalized optical power $I_{out}$', measured at one of the output ports of a MZI as a function of the electrical power $P$ dissipated by a microheater on a bridge waveguide: a complete reconfiguration is achieved with a power dissipation as low as 37 mW. The experimental dataset is reported along with its best sinusoidal fit, obtained by exploiting the mathematical model described by Equations 1 and 2.

### 3.2. Dynamic response

As shown in the previous section, thermal isolation has a beneficial effect on the static power dissipation $P_{2\pi}$. However, a higher isolation can slow down the dynamic response of the device and, thus, increase the switching time $\tau$ (10 to 90%) necessary for the reconfiguration. In order to take into account both the effects, silicon PICs are usually compared with the aid of a figure of merit (FOM) based on the product of these two quantities[7]. However, in order to compare devices operating at different wavelengths, it is possible to introduce a more general FOM, based on the former one and defined as

$$FOM = \frac{P_{2\pi}\tau}{\lambda}. \qquad (8)$$

Step response and FOM have been assessed by using the thermal shifter to induce a phase change $\Delta\varphi = \pi$ in a MZI, corresponding to a complete switch of the optical power from an output port to the other. The switching time $τ$ is reported in Figure 3a for the same MZIs presented in the previous section: isolation trenches slow down the step response, with a switching time that goes from 12 ms, for a MZI with no trenches, to 46 ms, for a MZI with a trench depth $D_t$ = 450 μm. However, if this worsening is considered along with the strong improvement in terms of static power dissipation, it is possible to conclude that the overall performance of the device benefits from the use of isolation trenches. This fact is quantified by the FOM, that is reported in Figure 3a along with the switching time: indeed, by introducing the isolation trenches the FOM improves (i.e. decreases) of about a factor of 3, starting from 4730 Ws/m and saturating down to about 1500 Ws/m. Again, it is worth noting that isolation trenches deeper than 300 μm do not provide any improvement.

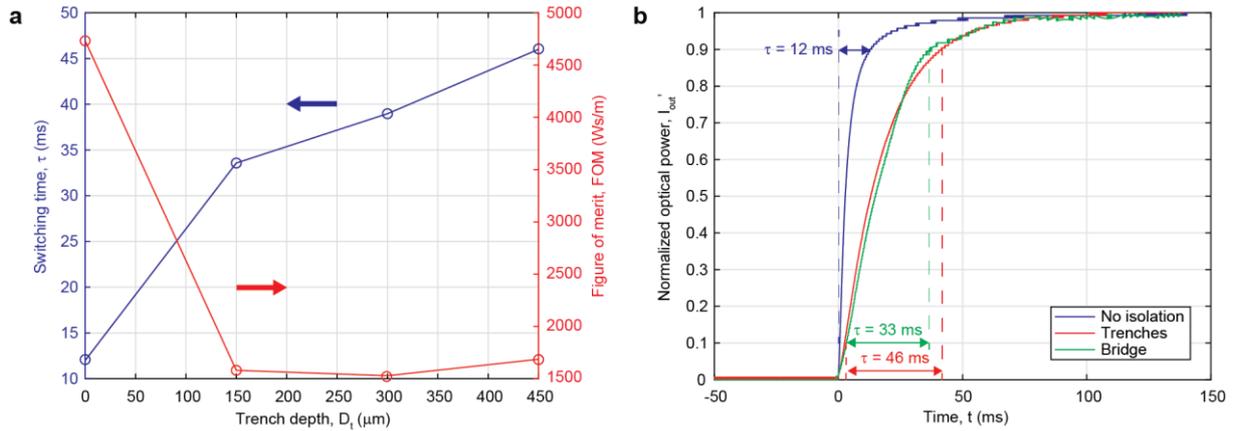

***Figure 3.*** *Dynamic response to a voltage step inducing a phase change Δφ = π. (a) Switching time τ and corresponding FOM as a function of the trench depth $D_t$. (b) Normalized optical power $I_{out}$' as a function of the time t for a MZI with no isolation, with isolation trenches ($D_t$ = 300 μm) and with a bridge waveguide.*

Finally, Figure 3b reports a comparison between the step responses, normalized to the final steady state value, for a MZI with no isolation, with isolation trenches ($D_t$ = 300 μm) and with a bridge waveguide. The latter features a switching time $τ$ = 33 ms, that is comparable to the one achieved with the trenches. Given the lower dissipated power, this result leads to a further improvement of the FOM that reaches a value as low as 788 Ws/m. Here we report only the heating transients, however it is worth stressing that both the heating and the cooling process have comparable switching time, as already reported for silicon circuits featuring suspended waveguides[31].

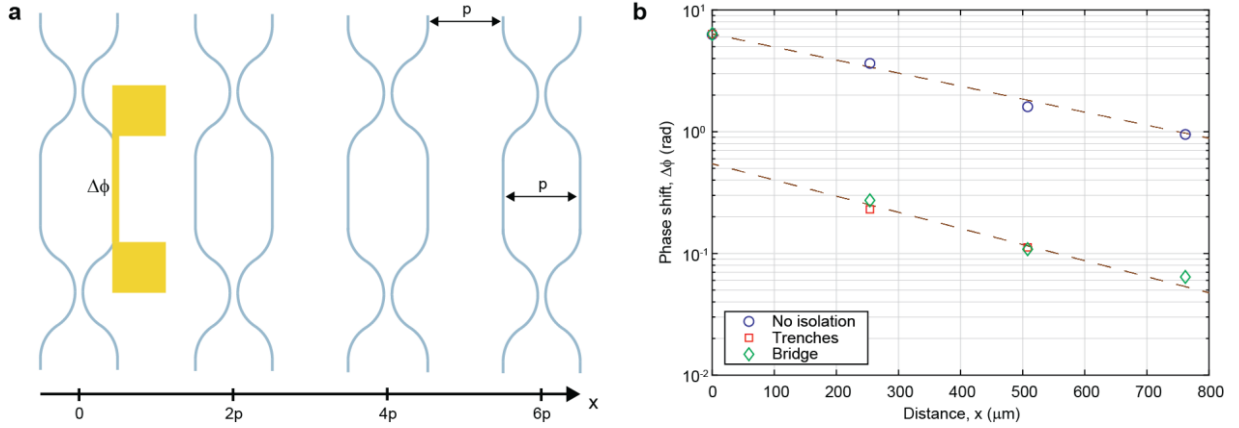

***Figure 4.*** *Thermal crosstalk induced by a thermal shifter on different MZIs. (a) Group of four MZIs used to characterize the phase induced by a shared microheater as a function of the distance x. (b) Phase shift Δφ as a function of the distance x between microheater and MZIs with no isolation, with isolation trenches ($D_t$ = 300 μm) and with a bridge waveguide. The inter-waveguide pitch is p = 127 μm. The dashed lines are guides for the eye.*

### 3.3. Thermal crosstalk

Up to now, we have considered a single MZI and the effects of a thermal shifter fabricated right upon it. However, due to thermal crosstalk, an interferometer can be affected also by the power dissipated in close-by devices, resulting in an undesired contribution $\Delta\varphi_{ct}$ to the actual phase $\varphi$. Under the linearity hypothesis stated in the former sections, the total phase $\varphi$ induced on a given MZI can be modeled by generalizing Equation 2 as

$$\varphi = \Phi + \Delta\varphi_0 + \Delta\varphi_{ct} = \Phi + \alpha_0 P_0 + \sum_{i=1}^{N-1} \alpha_i P_i, \qquad (9)$$

where $N$ is the total number of thermal shifters integrated in the device and the subscript 0 refers to the microheater right upon the MZI. In order to guarantee an effective control on the phase $\varphi$, the phase contribution $\Delta\varphi_{ct}$ induced by the other microheaters has to be as low as possible.

Thermal crosstalk has been measured on three groups of four MZIs each ($L$ = 3 mm, $p$ = 127 μm). In all the three groups there is one MZI with a microheater on top, each characterized by a different isolation approach (no isolation, isolation trenches having depth $D_t$ = 300 μm and the bridge waveguide). As depicted in Figure 4a, the four interferometers in each group have a different distance $x$ from the microheater, that ranges from $x$ = 0 μm (MZI fabricated beneath the thermal shifter) to $x = 6p$ = 762 μm (separation step $\Delta x = 2p$ = 254 μm). A comparison among the three isolation approaches is reported in Figure 4b in terms of the phase shift $\Delta\varphi$ as a function of the distance $x$. When the thermal shifter induces a $2\pi$ phase shift on the target interferometer (i.e. at $x = 0$ μm), the phase shift on the next-neighbor MZI (i.e. $x = 254$ μm) is $\Delta\varphi_{ct} = 3.64$ rad when no isolation is used. On the contrary, by exploiting one of the two isolation strategies, this value reduces to 0.23 rad for the trenches and to 0.27 rad for the bridge waveguide (corresponding to 3.7% and 4.3%, respectively, of the phase induced on the target). For MZIs that are further apart, the slope of the three curves is very similar, consistently with the fact that no isolation is replicated on these interferometers and, thus, all the improvement is gained in the isolation structure produced around the microheater.

### 3.4. Miniaturization

The reduction of the optical circuit dimensions would be advantageous not only to increase the integration density attainable with this platform, but also to reduce the insertion losses of the device as the propagation distance would be diminished. Firstly, it is interesting to investigate the scaling of the inter-waveguide distance $p$. A few problems can arise when reducing this parameter: indeed, by adopting $p = 80$ μm in interferometers that are not isolated, the power $P_{2\pi}$ dissipated by a thermal shifter set at $\Delta\varphi = 2\pi$ increases from 611 to 776 mW, while the phase $\Delta\varphi_{ct}$ induced by the same thermal shifter on an adjacent MZI (i.e. $x = 160$ μm) increases from 3.64 to 4.48 rad. On the contrary, the detrimental effects of the scaling are not observed on MZIs isolated with trenches ($D_t = 300$ μm): indeed, by adopting $p = 80$ μm on such interferometers, the power dissipation $P_{2\pi}$ remains as low as 57 mW, while the phase shift $\Delta\varphi_{ct}$ induced on the adjacent MZIs remains as low as 0.22 rad (3.5% of the phase induced on the target).

Secondly, another parameter that is desirable to scale is the thermal shifter length $L$. Since the thermal efficiency factor $\overline{R}$ does not depend on $L$, by considering Equation 4 one may think that this parameter can be reduced as much as we want, with no effect on the tuning performance of the circuit. Actually, this is not true: while the power dissipation $P$ needed to induce a given phase $\varphi$ is not dependent on $L$, the corresponding temperature difference $\Delta T$ induced between the two arms is inversely proportional to this parameter. Therefore, the microheater miniaturization leads to higher operating temperatures that, in turn, can degrade the resistive materials during the operation of the device. This phenomenon can lead to long-term drifts of the phase $\varphi$ or, in the worst case, to the breakdown of the thermal shifter. Mathematically speaking, Equation 5 can be exploited to estimate the maximum temperature difference $\Delta T_{2\pi}$ (i.e. the temperature difference corresponding to a $2\pi$ phase shift) as

$$\Delta T_{2\pi} = \frac{\lambda}{n_t L}. \qquad (10)$$

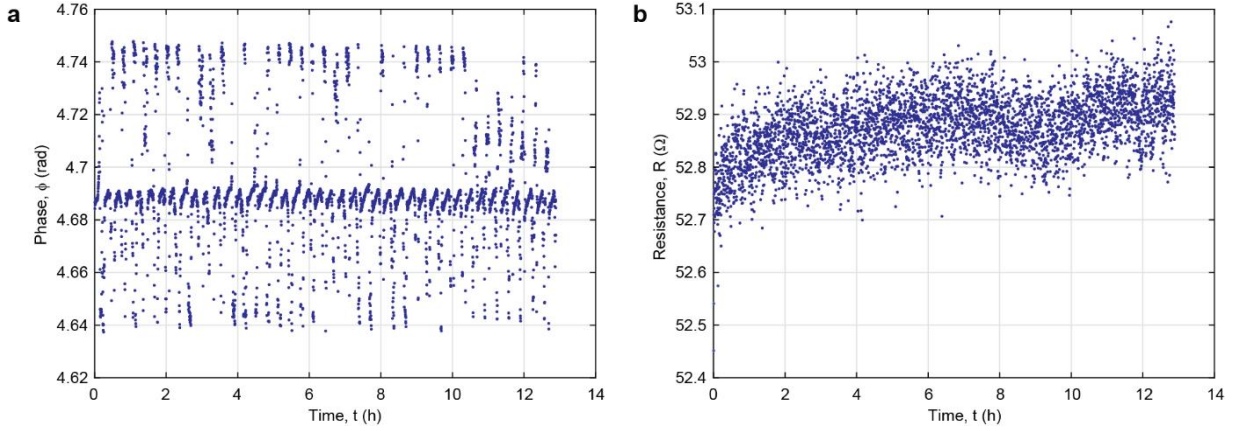

***Figure 5.*** *Stability of the thermal shifting operated on a bridge waveguide MZI (L = 1.5 mm). (a) Phase φ measured at the output of the MZI as a function of the time t. (b) Electrical resistance R of the microheater as a function of the time t.*

For the microheaters considered so far ($L$ = 3 mm) and a thermo-optic coefficient $n_t$ = 6.8E-6 K$^{-1}$, the temperature difference is $\Delta T_{2\pi}$ = 76 °C. The compatibility of this value with a stable operation of the thermal shifter has been already demonstrated[27], but it is interesting to investigate a further reduction of the length $L$. Therefore, we have characterized the phase stability of MZIs featuring 1.5 mm long microheaters on bridge waveguides, corresponding to an estimated temperature difference $\Delta T_{2\pi}$ = 152 °C. Figure 5a reports the phase $\varphi$ monitored from the output power distribution of such an interferometer, continuously operating for almost 13 h. The thermal shifter induces a constant phase $\varphi = \frac{3}{2}\pi$, a value that guarantees the maximum sensitivity of the measurement (see Equation 1) and, at the same time, a temperature difference close to the maximum one. No evidence of long-term drifts is present, while the short-term phase fluctuations are characterized by a small standard deviation of 23.5 mrad. Such fluctuation includes the effects strictly due to the thermal shifter, but also those due to the experimental setup (e.g. laser power fluctuations, alignment instabilities, room temperature changes, etc.). In order to isolate the contribution of the thermal shifter, the electrical resistance of the microheater has been monitored over the whole period of the measurement. Figure 5b reports this quantity as a function of the time: the standard deviation is as low as 59 m$\Omega$, corresponding to a phase variation of only 5.3 mrad. This is the value that could be achieved by the thermal tuning process if no other sources of fluctuations were present.

Finally, we investigated the electrical reliability of these devices on longer periods. To this aim, three microheaters have been continuously operated for over 2 weeks, with no evidence of damages or drifts of the electrical properties of the microheater.

### 3.5. Performance in vacuum

The experimental characterization presented so far demonstrates that the performance of these 3D-structured reconfigurable circuits is limited by the presence of air. Therefore, in order to investigate the possibility of further improving both the power dissipation and the thermal crosstalk, we have repeated in a vacuum chamber the main steps of the experimental characterization for a MZI ($L$ = 3 mm, $p$ = 127 μm) featuring a bridge waveguide. Thanks to a two stage pumping system, the experimental setup allows measurements at two different vacuum levels: a medium vacuum, corresponding to an absolute pressure P = 1.1E-4 bar, and a high vacuum, corresponding to P = 2.7E-7 bar. However, thanks to the fact that the pressure decreases with a slow transient, additional measurements can be collected at intermediate pressure values. In this way, the performance of the MZI can be characterized on an extended pressure range.

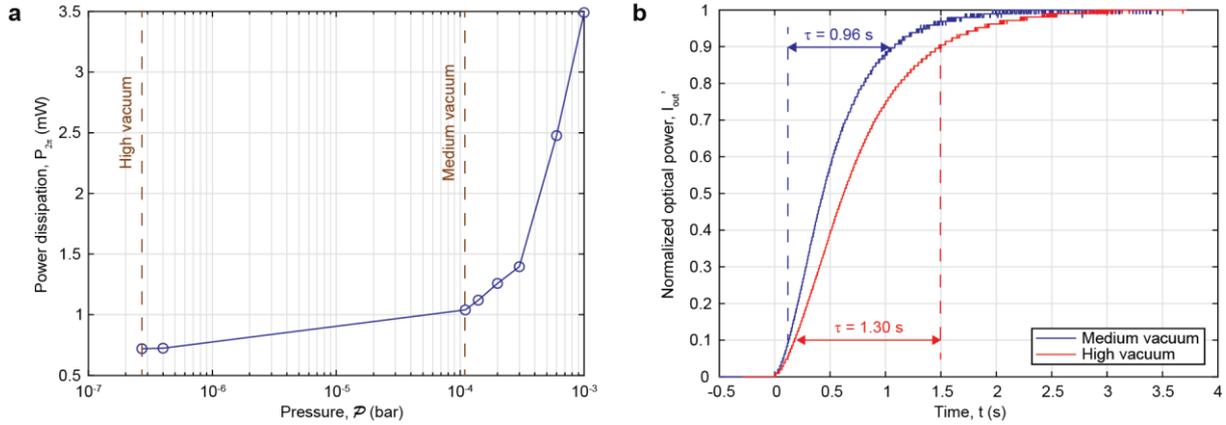

*Figure 6.* Experimental results obtained on a bridge waveguide MZI operated in vacuum. (a) Static characterization: power dissipation $P_{2\pi}$ as a function of the pressure P. (b) Step response characterization: normalized optical power $I_{out}'$ as a function of the time t for both the medium and the high vacuum levels achievable with the two stage pumping system.

This modus operandi is at the basis of Figure 6a, that reports the power dissipation $P_{2\pi}$ as a function of the pressure P: a value of 1E-3 bar is already enough to achieve $P_{2\pi} = 3.49$ mW, an improvement in terms of power dissipation of more than an order of magnitude with respect to the value measured in ambient conditions (i.e. P = 1 bar). By waiting for the pressure to reach the medium vacuum level, the value of $P_{2\pi}$ drops down to 1.04 mW. However, the power dissipation has not reached its minimum value yet: indeed, when the second pumping stage is turned on, this quantity further decreases and enters the sub-milliwatt range, with a power dissipation as low as 0.72 mW. Finally, it is worth noting that, close to the high vacuum level, the power dissipation $P_{2\pi}$ seems to become independent on the pressure P and, thus, it is possible to conclude that the minimum power dissipation achievable with the bridge waveguide is reached.

The excellent performance in terms of power dissipation is again counterbalanced by the slowing down of the step response. Figure 6b reports the normalized output power $I_{out}'$ as a function of the time *t*, when the thermal shifter is operated to induce a phase step having amplitude $\Delta\varphi = \pi$: the switching time increases from 33 ms to 0.96 s and 1.35 s for the medium and the high vacuum level. However, the overall performance of the device benefits from the operation in vacuum: indeed, the corresponding FOM drops to 644 and 604 Ws/m, respectively.

Finally, the experimental characterization in vacuum has been concluded with the assessment of the thermal crosstalk performance: when the microheater is biased in order to induce a 2π phase shift on the target MZI, the phase variation measured on the adjacent interferometer (i.e. $x$ = 254 μm) is as low as $\Delta\varphi_{ct}$ = 6 mrad (less than 0.1% of the phase induced on the target).

## 4. Conclusion

The technological platform that we have presented here demonstrates that it is possible to integrate state-of-the-art waveguide circuits inscribed by FLW with isolation structures that allow an efficient reconfiguration of the device. Indeed, the microstructures result in no compromise on the optical performance of the circuit, whose losses are comparable to the best results reported in the literature for FLW integrated circuits at telecom C-band wavelength[18,32]. In addition, by preserving the 3D capability and the polarization transparency of the circuits, the microstructures enhance the potentials of FLW without introducing substantial limitations. Two different isolation structures have been considered throughout the article: deep trenches and the bridge waveguide. Both provide a dramatic decrease in the dissipated power required to induce a 2π phase shift in a MZI, as well as in thermal crosstalk, with a better performance of the bridge waveguide. The improvement is also evident when both the static and the dynamic responses are assessed with the FOM that we have defined. A comparison among the best results achieved with FLW is reported in Table 1. The bridge waveguide compares favorably with the other FLW platforms, resulting the best solution both in terms of power dissipation and in terms of the overall performance, assessed by means of the FOM. Concerning the size of the FLW circuits, the devices presented in this work have been designed to achieve an integration scale that represents the state of the art for the FLW platform. With our 3D structuring method, we have demonstrated short microheaters, with length $L$ = 1.5 mm, providing an operation stability comparable to that achieved in longer FLW circuits[27], [28]. In addition, we have also demonstrated that a reduction of the inter-waveguide pitch, down to $p$ = 80 μm, can be achieved with negligible effect on both power dissipation and thermal crosstalk. These results pave the way towards a level of control, complexity and integration density never achieved before in a FLW device, opening exciting scenarios in photonic QIP.

| Work | Wavelength (nm) | Integration scale | | | Reconfigurability | | |
|---|---|---|---|---|---|---|---|
| | | Curvature radius (mm) | Inter-waveguide pitch (μm) | Microheater length (mm) | Power dissipation (mW) | Switching time (ms) | FOM (Ws/m) |
| Dyakonov et al.[26] | 800 | 80 | 100 | 3 | 336 - 1105 | 10 | 4200 - 13812 |
| Chaboyer et al.[28] | 800 | 50 | 254 | 12.5 | 200 | 400 | 100000 |
| This work - Bridge in air | 1550 | 45 | 127 | 3 | 37 | 33 | 788 |
| This work - Bridge at 1E-4 bar | 1550 | 45 | 127 | 3 | 1 | 960 | 644 |

*Table 1. Comparison among the best FLW platforms reported so far.*

A more general comparison can be made by considering also the reconfigurable PIC presented by Harris and co-workers[7], [8], which represents today the state of the art for the SOI platform. Regarding the optical circuit, the 8 dB overall losses of the silicon PIC are divided into two contributions: 1 dB on-chip losses (about 0.1 dB/interferometer) and 7 dB of coupling with the external fiber arrays. Our technology shows comparable results in terms of on-chip losses (about 0.35 dB/interferometer), but coupling losses are more than an order of magnitude lower (only 0.54 dB). Regarding the thermal phase shifters, it is the first time that a FLW device shows a comparable performance in terms of power dissipation and thermal crosstalk with respect to the best counterpart realized in a SOI platform. Indeed, even when operated in standard conditions (ambient pressure), our device shows a $P_{2\pi}$ = 37 mW (Harris et al.[7]: $P_{2\pi}$ = 49 mW) and 3.5% crosstalk (Harris et al.[8]: 1%). When operated in vacuum, the performance of our device becomes comparable ($P_{2\pi}$ = 1.04 mW and <0.1% crosstalk) with the one of the SOI suspended phase shifter of Fang and co-workers[31] ($P_{2\pi}$ = 0.98 mW). However, the remarkable results in terms of power dissipation and crosstalk come at a cost: the switching time of the circuit increases up to 960 ms. Nevertheless, for a large set of applications in QIP, once the target unitary operation is implemented, several data points need to be acquired in order to collect a significant statistics and even a switching time in the order of few seconds does not represent the bottleneck of the overall measurement time[33], [34].

All these results are even more remarkable if we think that they are demonstrated with relatively simple and inexpensive fabrication apparatuses. This is a not negligible advantage for the applications, if compared to the massive amount of resources that is needed to fabricate a PIC based on the SOI technology. In addition, by introducing trenches or bridges, the fabrication time of a single interferometer is increased up to only 15 minutes for a 1.5 mm long thermal phase shifter. This means that, by envisioning the fabrication of a complex PIC like the one reported by Harris et al.[7], [8], 176 phase shifters would require a total time of 44 h, still much shorter than the time required for a SOI circuit fabricated in a multi-project wafer[35], which is in the order of months, at the price of thousands of \$/mm$^2$. Finally, it is worth noting that cost, fabrication time and ease of handling will be moderately affected by the necessity of a vacuum environment. Indeed, some applications in QIP might require in any case the operation in vacuum[20] and, for all the other experiments, it is possible to encapsulate the device in order to create a localized vacuum around the circuit[36], a procedure considered today the routine in many integrated devices as inertial microelectromechanical systems (MEMS) or microbolometers. As a final remark, we would like to point out that reprogrammable quantum processors based on FLW can also operate in the visible and near-infrared wavelength range, which is an important requirement for multiple applications, from quantum memories[20] to quantum sensing of atomic transitions[37], free-space quantum key distribution[38], optical circuits coupled to quantum dot sources[39] and room-temperature quantum systems based on avalanche photodiodes instead of cryogenic single-photon detectors[40].

**Supporting Information**

Supporting Information is available from the Wiley Online Library or from the author.

**Acknowledgements**

This work has received funding from the European Research Council (ERC) under the European Union's Horizon 2020 research and innovation programme (project CAPABLE – Grant agreement No. 742745). A. C. and S. A. acknowledge support from the Italian Ministry of Education, University and Research (MIUR), PRIN 2017 programme, project ID 2017SRNBRK. The device fabrication was partially performed at PoliFAB, the micro- and nanofabrication facility of Politecnico di Milano (www.polifab.polimi.it). The authors would like to thank the PoliFAB staff (Claudio Somaschini, Alessia Romeo, Marco Asa and Lorenzo Livietti) for the valuable technical support. The authors would also like to thank Michele Devetta (IFN-CNR) and Salvatore Stagira (Dipartimento di Fisica - Politecnico di Milano) for providing the vacuum instrumentation and Gabriele Crippa (Dipartimento di Fisica - Politecnico di Milano) for the precious help during the vacuum characterization.

**Conflict of interest**

The authors declare no conflict of interest.